\documentclass[%
 reprint,
 amsmath,amssymb,
 aps,prl
]{revtex4-2}

\usepackage{graphicx}
\usepackage{dcolumn}
\usepackage{bm}
\usepackage{cleveref}

\usepackage{color}
\usepackage{mathrsfs}
\usepackage{here}

\begin{document}


\title{Solvable Neural Network Model for Input-Output Associations: Optimal Recall at the Onset of Chaos}

\author{Tomoki Kurikawa}
 \email{kurikawa@fun.ac.jp}
\affiliation{Department of Physics, Kansai Medical University, Shinmachi 2-5-1, Hirakata, Osaka, Japan}
\affiliation{Department of Complex and Intelligent systems, Future University Hakodate, 116-2 Kamedanakano-cho, Hakodate, Hokkaido, Japan 041-8655}

\author{Kunihiko Kaneko}%
\affiliation{
 The Niels Bohr Institute, University of Copenhagen, Blegdamsvej 17, Copenhagen, 2100-DK, Denmark
 }
\affiliation{
 Center for Complex Systems Biology, Universal Biology Institute, University of Tokyo, Komaba, Tokyo 153-8902, Japan
 }

\date{\today}

\begin{abstract}
    In neural information processing, an input modulates neural dynamics to generate a desired output.
    To unravel the dynamics and underlying neural connectivity enabling such input-output association, we proposed an exactly soluble neural-network model with a connectivity matrix explicitly consisting of inputs and required outputs. 
    An analytic form of the response upon the input is derived, whereas three distinctive types of responses including chaotic dynamics as bifurcation against input strength are obtained depending on the neural sensitivity and number of inputs. 
    Optimal performance is achieved at the onset of chaos, and the relevance of the results to cognitive dynamics is discussed.
\end{abstract}

\maketitle

Neural systems exhibit rich dynamics generated by strong recurrent connections\cite{Luczak2009}. 
For performing cognitive tasks in neural systems, sensory inputs modulate the neural dynamics to generate specific output patterns resulting in suitable behaviors.
In the association task between the input signals and output choices, for instance, the signal modifies ongoing (spontaneous) neural dynamics, leading to the emergence of an appropriate attractor that guides the correct choice\cite{Mante2013,Wang2018}, as strongly contrasted with input-output transformation in feed-forward networks\cite{Minsky, Aloysius2017}
Unveiling the mechanisms behind such modulation and the type of connectivity relevant to it is essential  for understanding information processing in neural systems. 

One widespread and powerful approach to understanding the information processing involves recurrent neural networks trained with machine learning techniques\cite{Mante2013,Barak2017,Chaisangmongkon2017,Wang2018,Kurikawa2018}. 
However, these trained networks are finely tuned for specific tasks, which masks the connectivity relevant to cognitive functions.
There is a need for a simple model to bridge the gap between neural connectivity and neural dynamics in shaping the input/output transformations.

Another approach, the auto-associative memory model, offers network connectivity explicitly represented by memorized patterns, as pioneered in the Hopfield network
\cite{Amari1977,Hopfield1984,Amit1987}.
In this approach, different fixed-point attractors correspond to distinct memorized patterns, to which neural states converge, depending on their initial states.
Thus, neural dynamics themselves are not modulated by the input.
The role of spontaneous dynamics without input and the connectivity underlying the change in the dynamics to produce the output remain to be elucidated.

In the present Letter, we propose a neural network model with a novel connectivity matrix designed to generate any memorized patterns when the associated inputs are applied with a certain strength.
This connectivity is explicitly represented based on a set of input and output patterns.
The recall to the input is given by the location of a fixed point for any strength of the input, which is analytically obtained in our model.
Besides this fixed-point, a chaotic attractor also emerges depending on the input strength, the gain parameter, and the number of memorized patterns.
We obtain the phase diagram on distinct recall behaviors against these parameters, which demonstrates that the highest performance in the recall is achieved at the onset of the chaos.
Finally, computational roles of chaotic internal dynamics are discussed in possible relation to experimental observations.

We consider a neural network model composed of $N$ neurons.
The network is required to generate target patterns $\bm{\xi^{\mu}}$ ($\mu=1,2,,,M$) in response to input patterns $\bm{\eta^{\mu}}$, where $M = \alpha N$, and, $\bm{\eta^{\mu}}$ and $\bm{\xi^{\mu}}$ are $N$-element vertical vectors.
Each element of these vectors takes a binary value ($\pm 1$) that is randomly generated according to the probability distribution $P(\xi_i^{\mu} = \pm 1) = P(\eta_i^{\mu} = \pm 1) = 1/2$.
The neural activity $x_i$ evolves according to the following equation:
\begin{align}
    \dot{x_{i}} =\tanh(\beta (\Sigma_j J_{ij} x_{j} + \gamma \eta_i^{\mu})) - x_{i},  \label{eq:neuro-dyn}
\end{align}
where $\beta$ and $\gamma$ are the gain of the activation function and the input strength, respectively.

To memorize input/output maps between $\boldsymbol{\eta}$ and $\boldsymbol{\xi}$, we have designed the connectivity matrix  $J$ that is composed of $\boldsymbol{\eta}$ and $\boldsymbol{\xi}$, in contrast to the Hopfield network that incorporates only $\boldsymbol{\xi}$.
Further, to mitigate the effects of potential interference across memories that could impair recall performance\cite{Personnaz1986,Kanter1987,Diederich1987}, the designed connectivity is given with a pseudo-inverse matrix of the target-input matrix $X$  as follows:

\begin{align}
    J &=& X \left(
    \begin{array}{cc}
        I & I \\
        -I & -I
    \end{array}
    \right) X^{+} \\
    X &=& [\bm{\xi^{1}},\boldsymbol{\xi^2},\ldots,\bm{\xi^{M}},\bm{\eta^1},\boldsymbol{\eta^2},\ldots,\bm{\eta^M}],\label{eq:Jstructure}    
\end{align}
where $I$ is an $M$-dimensional identity matrix, $X$ is an $(N,2M)$-matrix and $X^{+} \triangleq (X^{T}X)^{-1}X^{T}$ is a pseudo-inverse matrix of $X$, where $X^{T}$ is a transpose matrix of $X$.
Due to the pseudo-inverse matrix, $J\boldsymbol{\xi}^{\mu} + \gamma \boldsymbol{\eta^{\mu}} = \boldsymbol{\xi^{\mu}} +(\gamma-1)\boldsymbol{\eta^{\mu}} $ and, consequently, the target $\boldsymbol{\xi^{\mu}}$ is a fixed point under $\boldsymbol{\eta^{\mu}}$ with $\gamma=1$ for $\beta \rightarrow \infty$,  based on the properties of $\tanh(\beta x)$.
This property applies to all  $\mu$, indicating that all $\boldsymbol{\xi}^{\mu}$ are the fixed points under the corresponding inputs with $\gamma=1$.
In other words, all associations are successfully memorized in this model.
To satisfy the pseudo-inverse matrix, however, 
the number of vectors, $2M$, that are linearly independent of each other should be less than $N$.
As a consequence, at best, $M=N/2$ associations are allowed and the memory capacity is bounded by $\alpha = 0.5$ at a maximum.

\begin{figure}[t]
  \begin{center}
    \includegraphics[width=50mm]{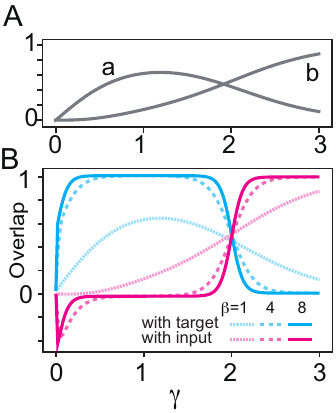}
    \caption { 
        Analytically obtained response of the network to input with increasing the input strength $\gamma$.
        A) $a(\gamma)$ and $b(\gamma)$ in Eqs.\ref{eq:ab1} and \ref{eq:ab2}, for $\beta=1$.
        B) The overlaps of $x^{\text{fp}}$ with a target and an input for increasing $\gamma$, plotted for different $\beta$ in blue and red, respectively.
        }
    \label{fig:rcl_reg}
  \end{center}
\end{figure}

How does the network recall to the input except for $\gamma = 1$ and $\beta \rightarrow \infty$?
We, now, derive an analytical form of a fixed point of the neural dynamics upon input for any value of $\gamma$ with finite $\beta$.
For it, we consider $x^{\text{fp}}(\gamma)=(a(\gamma) \bm{\xi} + b(\gamma) \bm{\eta})$ and derive $a(\gamma)$ and $b(\gamma)$ such that satisfy the fixed point condition for any $\gamma$ as follows.
Below, the superscript $\mu$ is omitted for clarity unless otherwise noted since the result is not dependent on $\mu$.
By using $J\boldsymbol{\xi}=J\boldsymbol{\eta}=\boldsymbol{\xi}-\boldsymbol{\eta}$, we have 
\begin{align}
    Jx^{\text{fp}}  = (a+b)(\boldsymbol{\xi}-\boldsymbol{\eta}),
    \label{eq:J_trans}
\end{align}
and, subsequently, by substituting $x^{\text{fp}}$ to $\dot{x}=0$ in Eq. \ref{eq:neuro-dyn},
\begin{align}
    a\boldsymbol{\xi}+b\boldsymbol{\eta}=f((a+b)(\boldsymbol{\xi}-\boldsymbol{\eta}) + \gamma \boldsymbol{\eta}),
\end{align}
where $f(x)=\tanh(\beta x)$.
Considering $i$-th elements such that $\xi_i$ equals $\eta_i$,
$a+b=f(\gamma)$ should be satisfied and, similarly, by considering $i$-th elements such that $\xi_i$ equals $-\eta_i$, $a-b=f(2(a+b)-\gamma)$ should be satisfied.
Thus we derive $a$ and $b$ as
\begin{align}
    a=(f(\gamma)+f(2f(\gamma)-\gamma))/2,  \label{eq:ab1} \\
    b=(f(\gamma)-f(2f(\gamma)-\gamma))/2, \label{eq:ab2}
\end{align}
where $a$ and $b$ are uniquely determined and depend solely on $\gamma$ for a given activation function $f(x)$
while they are independent of $N$, $\alpha$.
It is straightforward to check that Eq.\ref{eq:J_trans} is satisfied for any $\boldsymbol{\xi}$ and $\boldsymbol{\eta}$.
Although not proven analytically, we have confirmed numerically that $x^{\text{fp}}$ is a unique fixed-point for given $\mu$ and $\gamma$.
This rigorous solution is applicable not only to $\tanh(\cdot)$ but also to any arbitrary function, as long as $\boldsymbol{\xi}$ and $\boldsymbol{\eta}$ are binary vectors.
As $\gamma$ increases from zero, $a(\gamma)$ increases and takes a peak for $\gamma=1$, while $b(\gamma)$ increases more slowly as plotted in Fig.\ref{fig:rcl_reg}A.
For $\gamma$ less than $2$, $a(\gamma)$ is larger than $b(\gamma)$ and, oppositely, beyond $\gamma = 2$, $b(\gamma)$ is larger than $a(\gamma)$\footnote{When $\beta \rightarrow \infty$, $a=1$ and $b=0$ for $0 < \gamma <2 $, while $a=0$ and $b=1$ for $2 < \gamma $, which are obtained from Eqs. \ref{eq:ab1} and \ref{eq:ab2} }.
The overlap of $x^{\text{fp}}$ with $\boldsymbol{\xi}$,  termed $m \triangleq \Sigma_i x_i^{\text{fp}} \xi_i/N$, is also plotted in Fig. \ref{fig:rcl_reg}B. 
For a given $\beta$, $m$ increases up to $\gamma=1$ and, subsequently, decreases.
As $\beta$ increases, the curve of $m$ is steeper so that $x^{\text{fp}}$ nearly equals 1 even for the weak input.
The overlap of $x^{\text{fp}}$ with $\boldsymbol{\eta}$  slowly increases with $\gamma$, followed by a sharp rise at $\gamma\approx 2$ beyond which it approaches unity, i.e., the network just outputs the input as it is (Fig. \ref{fig:rcl_reg}B).
Thus, in the following part, we consider the range of $0 \leq \gamma \leq 2$.

\begin{figure*}[t]
  \begin{center}
    \includegraphics[width=140mm]{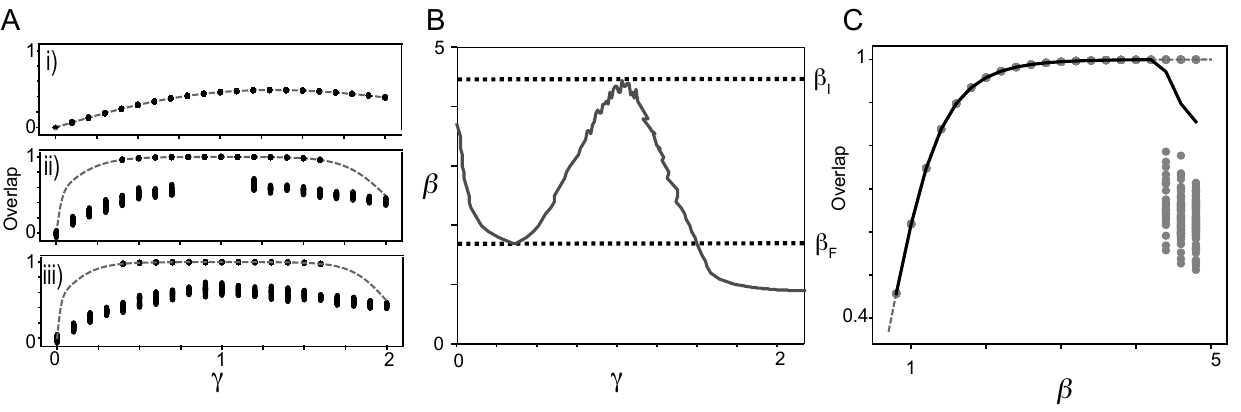}
    \caption { 
    Three recall behaviors in response to $\boldsymbol{\eta}$ depending on $\beta$.
        A) The overlaps $m$ against the increase in $\gamma$  are shown for $\beta=0.8,4,4.7$ in (i)-(iii) panels, respectively. 
        In each panel, each of the black dots for a given $\gamma$ represents the overlap of $x$ with $\boldsymbol{\xi}^1$ averaged over 100 unit-time after the transient period.
        To confirm the stability of $x^{\text{fp}}$ and explore another attractor, we sampled the dynamics from 20 random initial states in addition to an initial state equal to $x^{\text{fp}}$.
        The dotted lines represent the overlap of $x^{\text{fp}}$ with $\boldsymbol{\xi}^1$ as also shown in panel C.
        B) Stability of the chaotic attractor against $\beta$ and $\gamma$. 
        The chaotic attractor is present above the curve.
        All results are obtained for $\alpha=0.38$.
        C) The overlap at $\gamma=1$ with the increase in $\beta$, as in A.
        Dots represent the overlaps obtained from 100 randomly chosen initial states, while the solid line exhibits the overlap averaged over them.
        }
    \label{fig:recall}
  \end{center}
\end{figure*}

\begin{figure*}[t]
  \begin{center}
    \includegraphics[width=160mm]{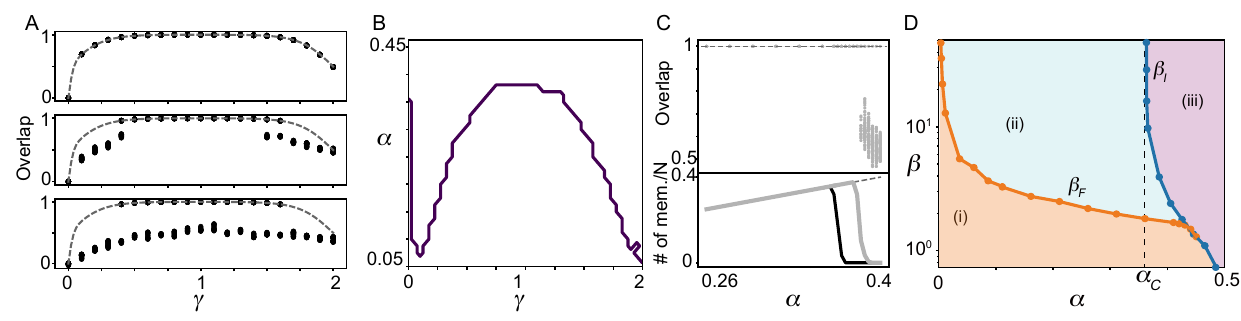}
    \caption { 
    Three recall behaviors in response to $\boldsymbol{\eta}$ depending on $\alpha$.
        A) The overlaps against the increase in $\gamma$ are shown for $\alpha=0.05,0.3,0.4$ in (i)-(iii) panels, respectively. 
        Black dots and dotted lines exhibit the neural states and  $x^{\text{fp}}$  in the same way as in Fig. \ref{fig:recall}A. 
        B) Stability of the chaotic attractor against $\alpha$ and $\gamma$. 
        The chaotic attractor is present above the red curve.
        All results are obtained for $\beta=4$.
        C) (Upper panel) Bifurcation diagram of the overlap at $\gamma=1$ with the increase in $\alpha$ is shown in the same way as in A.
       (Lower panel) The number of memorized patterns (i.e., the number of  $x^{\text{fp},\mu}$ ($\mu = 1, \ldots  ,\alpha N$ ) into which the neural state converges) normalized by $N$ is plotted. Filled lines in  gray  and black show the behavior for $\beta=4,32$, respectively. 
        The dotted line represents the maximum number of possible memories normalized by $N$ (i.e., $\alpha$).
        D)    Three recall behaviors in response to $\boldsymbol{\eta}$.
    A phase diagram of the recall regimes (i,ii,iii) against $\alpha$ and $\beta$.
 $\beta_F$ (orange) gives the boundary of the stable $\boldsymbol{x}^{\text{fp}}$,
while $\beta_I$ (magenta) shows the border of the impaired recall regime.
        }
    \label{fig:depend_a}
  \end{center}
\end{figure*}

Although $x^{\text{fp}}$ is a fixed point for any value of parameters, it is necessary to ascertain its stability and the existence of other attractors, to assess whether the recall of $x^{\text{fp}}$ really works from any initial states.
We numerically solved Eq. (1)  ($N=2048$, unless otherwise noted),
and found another chaotic attractor in addition to $x^{\text{fp}}$.
By varying $\alpha$ and $\beta$, three types of recall behaviors are observed depending on the stability of these two attractors, which are characterized by the distinct bifurcation diagrams of $m$ against $\gamma$  (as shown in Fig. \ref{fig:recall}A(i)-(iii)):
(i) Stable recall of $x^{\text{fp}}$ for any strength of the input: $x^{\text{fp}}$ is a unique attractor for any $\gamma$. 
(ii) Stable recall of $x^{\text{fp}}$ only for a certain range of $\gamma$: $x^{\text{fp}}$ is a unique attractor for $\gamma \sim 1$, whereas 
for smaller $\gamma$ the chaotic attractor appears, which exhibits a smaller overlap with $\xi$ compared with the overlap of $x^{\text{fp}}$\footnote{$x^{\text{fp}}$ and the chaotic attractor coexist for some range of parameters and do not otherwise. Even when they coexist, the neural states converge into the chaotic attractor for almost all initial states, as shown in Figs. \ref{fig:recall}C, \ref{fig:depend_a}C and S1C. Thus, the type of recall is independent of their coexistence.}.
For smaller $\gamma$ values, the neural state fails to converge into $x^{\text{fp}}$, and instead, it converges into the chaotic attractor from most initial states, meaning that the network fails to recall the target.
Still,  for $\gamma \sim 1$, the neural state from any initial state converges to $x^{\text{fp}}$ whose overlap with the target is  close to unity, resulting in the recall of the target.
(iii) No stable recall of $x^{\text{fp}}$ for any $\gamma$: the chaotic attractor exists across all ranges of $\gamma$, 
even though $x^{\text{fp}}$ coexists around $\gamma=1$.
The chaotic attractor has a much larger basin of attraction than $x^{\text{fp}}$ even for $\gamma \sim 1$ (Fig. \ref{fig:dyn_depn_b}C).
Consequently, the recall of the target is impaired.

To analyze these three behaviors, we first explored the stability of $x^{\text{fp}}$ and of the chaotic attractor across a range of  $\beta$ with a constant $\alpha=0.38$. 
We found that for a small value of  $\beta$ ($\beta=0.8$), the stable recall (i) is achieved.
The neural states from any initial states for any $\gamma$ converge rapidly into $x^{\text{fp}}$ (as shown in Fig. \ref{fig:dyn_depn_b}A),
indicating high robustness in the success recall.
However, the degree of overlap with the target is notably below the unity.

As $\beta$ increases, $x^{\text{fp}}$ approaches the target for all ranges of $\gamma$.
Beyond the critical $\beta$, denoted by $\beta_F$, $x^{\text{fp}}$ turns to be unstable for a certain range of $\gamma$, while the chaotic attractor emerges, corresponding to the recall type (ii) as shown in Fig. \ref{fig:recall}A(ii).
The overlap of the chaotic attractor with  the target is much lower than that of $x^{\text{fp}}$.
Although, for $\gamma=1$,  $x^{\text{fp}}$ is the unique attractor, there exists long-term transient chaos before the neural state converges into $x^{\text{fp}}$ (see Fig. \ref{fig:dyn_depn_b}B).

With the further increase in $\beta$,  the range of $\gamma$ within which the chaotic attractor exists expands, eventually, covering  all the range of $0 \leq \gamma \leq 2$ at another critical value of $\beta$ (termed $\beta_I$). 
Beyond $\beta_I$, the system exhibits the recall type (iii).
Even for $\gamma=1$, the basin of the chaotic attractor covers the full state space, and most orbits from random initial conditions converge into it (Fig. \ref{fig:dyn_depn_b}C).
Thus, the recall of the target almost fails.

To comprehensively understand the  recall behavior across $\beta$ and $\gamma$, we draw the regions where the chaotic attractor is present in Fig. \ref{fig:recall}B.
We also investigated the stability of $x^{\text{fp}}$, which is, however, not directly related to the type of recall and is shown in Fig. \ref{fig:stablity_xfp}.
In the area above the curve, the chaotic attractor is present.
$\beta_F$ is the minimum value of the curve in $\beta \leq 1$,
whereas $\beta_I$ is the maximum value of $\beta$ on the curve at $\gamma=1$.
These two critical values of $\beta$ determine the phase boundary of three recall behaviors (i)-(iii).

With an increase in $\beta$, $x^{\text{fp}}$ approaches the target, and accordingly, the final states in all the recall trials overlap almost perfectly with the target below  $\beta=\beta_I$ at which the chaotic attractor emerges (Fig. \ref{fig:recall}C).
As $\beta$ increases beyond $\beta_I$, the basin of the $x^{\text{fp}}$ attractor shrinks, while that of the chaotic attractor expands.
Consequently, the overlap between the final state and the target averaged over randomly chosen initial states significantly decreases, as depicted in Fig. \ref{fig:recall}C. 
Thus, the recall performance reaches its peak (i.e., at the onset of chaos) across all ranges of $\gamma$.

So far, we presented the results with the fixed number of memories $\alpha N$ ($\alpha=0.38$).
Noting that standard associative memory models such as the Hopfield network, recall fails beyond a critical number of embedded memories. 
We next analyze the change in the recall process with increasing $\alpha$ and demonstrate it exhibits similar behavior to the change with the increases in $\beta$:
Three types of recall behavior emerge as $\alpha$ varies, as shown in Fig. \ref{fig:depend_a}A.
For small $\alpha$, $x^{\text{fp}}$ is stable and a unique attractor for any $\gamma$ (type (i) ).
With the increase in $\alpha$, the chaotic attractor emerges within a certain range of $\gamma$   (type (ii) ), and this range expands (Fig.\ref{fig:depend_a}B).
Finally, the range within which the chaotic attractor is present covers all the ranges of $\gamma$  (type (iii) ).
In contrast to the clear change in the recall process with increasing $\beta$, the value of $x^{\text{fp}}$ remains  unchanged during the increase in $\alpha$.

We now focus on the behavior for $\gamma=1$ and explore the number of memories recalled successfully.
We found that at $\alpha = \alpha_C (\beta) $ the chaotic attractors emerge for all embedded patterns $\mu$ ($\alpha_C(\beta)$ is obtained by solving $\beta=\beta_I (\alpha)$).
For $\alpha < \alpha_C$, the fixed points $x^{\text{fp},\mu}=(a \bm{\xi}^{\mu} + b \bm{\eta}^{\mu})$ for all $\mu$ are stable and all the embedded patterns are successfully recalled, 
whereas for $\alpha>\alpha_C$, almost all the recall trials fail for all patterns due to the emergence of the chaotic attractors whose basins of attraction are much larger than those of $x^{\text{fp},\mu}$.
The number of successfully recalled memories increases linearly below $\alpha=\alpha_C(\beta)$ and then drops to zero drastically (see Fig. \ref{fig:depend_a}C,) signifying that $\alpha_C(\beta) N$ is the memory capacity in this model (e.g., $\alpha_C (4) = 0.38$).
$\alpha_C (\beta)$ decreases towards a certain finite value $\alpha_C(\infty)$ with the increase in $\beta$ as analyzed in detail in the following.

We finally show the phase diagram of the recall process against $\alpha$ and $\beta$ by identifying $\beta_F(\alpha)$ and $\beta_I(\alpha)$  ($\alpha_C (\beta)$ is the inverse function of $\beta_I (\alpha)$) as shown in Fig \ref{fig:depend_a}D.
As $\alpha$ approaches zero, $\beta_F$ diverges, meaning that if $\alpha$ is set to a sufficiently small value, $x^{\text{fp}}$ is stable throughout all $\gamma$ even for quite large $\beta$.
For such a limit, $x^{\text{fp}}$ approaches a step function;  $x^{\text{fp}}=1$ for $0 < \gamma < 2$ and $x^{\text{fp}}=0$ for otherwise.
Consequently, the network perfectly recalls the target for $0 < \gamma < 2$.

$\beta_I$ increases drastically as $\alpha$ decreases from $0.5$ and diverges at $\alpha_C(\infty)$.
For $\alpha$ below $\alpha_C (\infty)$,
the neural state converges to $x^{\text{fp}}$ for $\gamma=1$ even for $\beta \rightarrow \infty$.
The asymptotic analysis demonstrates that $\alpha_C(\infty) \sim 0.340$ for $N \rightarrow \infty$ (See Fig. \ref{fig:alpha_c}), 
indicating that the memory capacity is $\alpha = 0.340$ when $\beta$ is sufficiently large.

In summary, we present an analytically solvable neural network model for I/O associations in which each input stabilize the target pattern as a unique fixed-point under the limit of memory capacity.
This connectivity in the network consists of both target and input patterns, by introducing the pseudo-inverse matrix, which allows for rigorous recalls of any (correlated) target patterns.
This is in contrast to our previous model\cite{Kurikawa2012} valid only for mutually orthogonalized patterns.
By using this model, we derive the response to the input as the analytical expression of the fixed point for any input strength, whereas the response dynamics were explored in random networks (without embedded patterns)\cite{Rajan2010,Toyoizumi2011,Takasu2023} and low-rank networks\cite{Mastrogiuseppe2018,Schuessler2020}.
We also demonstrate the emergence of the additional chaotic attractor numerically.
Through exploration of the stability of these attractors, we identified three distinct recall processes.

Introducing the pseudo-inverse matrix ($X^{+}$  in Eq. \ref{eq:Jstructure}) into the connectivity generally requires the global information of the network, which may be difficult to implement biologically( but see\cite{Diederich1987} for Hopfield network).
In our previous study\cite{Kurikawa2013,Kurikawa2020}, however, a Hebbian and anti-Hebbian learning that only requires local information can shape the connectivity that is similar to our current connectivity.
Still, filling the gap between the learning-shaped connectivity and the current connectivity needs further studies.

Here, we uncovered three phases of recalls, concerning the dominance of the chaotic attractor.
Interestingly, the recall performance is maximized at the onset of the chaos, where the spontaneous chaotic activity is bifurcated to the fixed point that corresponds to the target output. 
In fact, such transitions of the activities with changes in the stimuli are observed in many cortical areas\cite{Churchland2010,Niessing2010}
These are consistent with our findings of the optimal performance under spontaneous chaotic dynamics, 
whereas the roles of the chaotic dynamics in the response and learning need to be further elucidated.
Indeed, the relevance of spontaneous chaotic (and high-dimensional) dynamics to computational neuroscience has been discussed, for instance, in the reservoir computing\cite{Maass2002,Jaeger2004,Bertschinger2004,Legenstein2007}, memories\cite{Tsuda2001}, mixed-selectivity for efficient separation\cite{Fusi2016}, sampling\cite{Terada2023}, neural avalanche\cite{Petermann2009,Kusmierz2020} and learning\cite{Sussillo2009, Orhan2019}.
Our study has demonstrated a new role of chaotic dynamics in recall performance.

Although Hopfield networks\cite{Amari1977,Hopfield1984} and their variants\cite{Personnaz1986,Kanter1987,Diederich1987} 
have great contributions to associative memory, 
the modulation of the internal dynamics by external input that is essential for performing cognitive functions has not been included.
Our model presents a novel prototype connectivity underlying such modulation, which will advance our understanding of neural processing.

\section{Acknowledgments}
T.K. and K.K. are supported by JSPS KAKENHI (No.20H00123, T.T and K.K) and Novo Nordisk Foundation (0065542, K.K)

\bibliography{11th_refs}

\begin{thebibliography}{37}%
\makeatletter
\providecommand \@ifxundefined [1]{%
 \@ifx{#1\undefined}
}%
\providecommand \@ifnum [1]{%
 \ifnum #1\expandafter \@firstoftwo
 \else \expandafter \@secondoftwo
 \fi
}%
\providecommand \@ifx [1]{%
 \ifx #1\expandafter \@firstoftwo
 \else \expandafter \@secondoftwo
 \fi
}%
\providecommand \natexlab [1]{#1}%
\providecommand \enquote  [1]{``#1''}%
\providecommand \bibnamefont  [1]{#1}%
\providecommand \bibfnamefont [1]{#1}%
\providecommand \citenamefont [1]{#1}%
\providecommand \href@noop [0]{\@secondoftwo}%
\providecommand \href [0]{\begingroup \@sanitize@url \@href}%
\providecommand \@href[1]{\@@startlink{#1}\@@href}%
\providecommand \@@href[1]{\endgroup#1\@@endlink}%
\providecommand \@sanitize@url [0]{\catcode `\\12\catcode `\$12\catcode
  `\&12\catcode `\#12\catcode `\^12\catcode `\_12\catcode `\%12\relax}%
\providecommand \@@startlink[1]{}%
\providecommand \@@endlink[0]{}%
\providecommand \url  [0]{\begingroup\@sanitize@url \@url }%
\providecommand \@url [1]{\endgroup\@href {#1}{\urlprefix }}%
\providecommand \urlprefix  [0]{URL }%
\providecommand \Eprint [0]{\href }%
\providecommand \doibase [0]{https://doi.org/}%
\providecommand \selectlanguage [0]{\@gobble}%
\providecommand \bibinfo  [0]{\@secondoftwo}%
\providecommand \bibfield  [0]{\@secondoftwo}%
\providecommand \translation [1]{[#1]}%
\providecommand \BibitemOpen [0]{}%
\providecommand \bibitemStop [0]{}%
\providecommand \bibitemNoStop [0]{.\EOS\space}%
\providecommand \EOS [0]{\spacefactor3000\relax}%
\providecommand \BibitemShut  [1]{\csname bibitem#1\endcsname}%
\let\auto@bib@innerbib\@empty
\bibitem [{\citenamefont {Luczak}\ \emph {et~al.}(2009)\citenamefont {Luczak},
  \citenamefont {Bartho},\ and\ \citenamefont {Harris}}]{Luczak2009}%
  \BibitemOpen
  \bibfield  {author} {\bibinfo {author} {\bibfnamefont {A.}~\bibnamefont
  {Luczak}}, \bibinfo {author} {\bibfnamefont {P.}~\bibnamefont {Bartho}},\
  and\ \bibinfo {author} {\bibfnamefont {K.~D.}\ \bibnamefont {Harris}},\
  }\bibfield  {title} {\bibinfo {title} {{Spontaneous Events Outline the Realm
  of Possible Sensory Responses in Neocortical Populations}},\ }\href
  {http://www.sciencedirect.com/science/article/B6WSS-4W8VHPH-D/2/6bc6e52b812b23cbb64603f59c46e72b}
  {\bibfield  {journal} {\bibinfo  {journal} {Neuron}\ }\textbf {\bibinfo
  {volume} {62}},\ \bibinfo {pages} {413} (\bibinfo {year} {2009})}\BibitemShut
  {NoStop}%
\bibitem [{\citenamefont {Mante}\ \emph {et~al.}(2013)\citenamefont {Mante},
  \citenamefont {Sussillo}, \citenamefont {Shenoy},\ and\ \citenamefont
  {Newsome}}]{Mante2013}%
  \BibitemOpen
  \bibfield  {author} {\bibinfo {author} {\bibfnamefont {V.}~\bibnamefont
  {Mante}}, \bibinfo {author} {\bibfnamefont {D.}~\bibnamefont {Sussillo}},
  \bibinfo {author} {\bibfnamefont {K.~V.}\ \bibnamefont {Shenoy}},\ and\
  \bibinfo {author} {\bibfnamefont {W.~T.}\ \bibnamefont {Newsome}},\
  }\bibfield  {title} {\bibinfo {title} {{Context-dependent computation by
  recurrent dynamics in prefrontal cortex.}},\ }\href
  {https://doi.org/10.1038/nature12742} {\bibfield  {journal} {\bibinfo
  {journal} {Nature}\ }\textbf {\bibinfo {volume} {503}},\ \bibinfo {pages}
  {78} (\bibinfo {year} {2013})}\BibitemShut {NoStop}%
\bibitem [{\citenamefont {Wang}\ \emph {et~al.}(2018)\citenamefont {Wang},
  \citenamefont {Narain}, \citenamefont {Hosseini},\ and\ \citenamefont
  {Jazayeri}}]{Wang2018}%
  \BibitemOpen
  \bibfield  {author} {\bibinfo {author} {\bibfnamefont {J.}~\bibnamefont
  {Wang}}, \bibinfo {author} {\bibfnamefont {D.}~\bibnamefont {Narain}},
  \bibinfo {author} {\bibfnamefont {E.~A.}\ \bibnamefont {Hosseini}},\ and\
  \bibinfo {author} {\bibfnamefont {M.}~\bibnamefont {Jazayeri}},\ }\bibfield
  {title} {\bibinfo {title} {{Flexible timing by temporal scaling of cortical
  responses}},\ }\href {https://doi.org/10.1038/s41593-017-0028-6} {\bibfield
  {journal} {\bibinfo  {journal} {Nature Neuroscience}\ }\textbf {\bibinfo
  {volume} {21}},\ \bibinfo {pages} {102} (\bibinfo {year} {2018})}\BibitemShut
  {NoStop}%
\bibitem [{\citenamefont {Minsky}\ and\ \citenamefont
  {Seymour}(1969)}]{Minsky}%
  \BibitemOpen
  \bibfield  {author} {\bibinfo {author} {\bibfnamefont {M.}~\bibnamefont
  {Minsky}}\ and\ \bibinfo {author} {\bibfnamefont {P.}~\bibnamefont
  {Seymour}},\ }\href@noop {} {\emph {\bibinfo {title} {{Perceptrons.}}}}\
  (\bibinfo  {publisher} {M.I.T. Press},\ \bibinfo {year} {1969})\BibitemShut
  {NoStop}%
\bibitem [{\citenamefont {Aloysius}\ and\ \citenamefont
  {Geetha}(2017)}]{Aloysius2017}%
  \BibitemOpen
  \bibfield  {author} {\bibinfo {author} {\bibfnamefont {N.}~\bibnamefont
  {Aloysius}}\ and\ \bibinfo {author} {\bibfnamefont {M.}~\bibnamefont
  {Geetha}},\ }\bibfield  {title} {\bibinfo {title} {{A review on deep
  convolutional neural networks}},\ }in\ \href
  {https://doi.org/10.1109/ICCSP.2017.8286426} {\emph {\bibinfo {booktitle}
  {2017 International Conference on Communication and Signal Processing
  (ICCSP)}}}\ (\bibinfo  {publisher} {IEEE},\ \bibinfo {year} {2017})\ pp.\
  \bibinfo {pages} {0588--0592}\BibitemShut {NoStop}%
\bibitem [{\citenamefont {Barak}(2017)}]{Barak2017}%
  \BibitemOpen
  \bibfield  {author} {\bibinfo {author} {\bibfnamefont {O.}~\bibnamefont
  {Barak}},\ }\bibfield  {title} {\bibinfo {title} {{Recurrent neural networks
  as versatile tools of neuroscience research}},\ }\href
  {https://doi.org/10.1016/j.conb.2017.06.003} {\bibfield  {journal} {\bibinfo
  {journal} {Current Opinion in Neurobiology}\ }\textbf {\bibinfo {volume}
  {46}},\ \bibinfo {pages} {1} (\bibinfo {year} {2017})}\BibitemShut {NoStop}%
\bibitem [{\citenamefont {Chaisangmongkon}\ \emph {et~al.}(2017)\citenamefont
  {Chaisangmongkon}, \citenamefont {Swaminathan}, \citenamefont {Freedman},\
  and\ \citenamefont {Wang}}]{Chaisangmongkon2017}%
  \BibitemOpen
  \bibfield  {author} {\bibinfo {author} {\bibfnamefont {W.}~\bibnamefont
  {Chaisangmongkon}}, \bibinfo {author} {\bibfnamefont {S.~K.}\ \bibnamefont
  {Swaminathan}}, \bibinfo {author} {\bibfnamefont {D.~J.}\ \bibnamefont
  {Freedman}},\ and\ \bibinfo {author} {\bibfnamefont {X.-J.}\ \bibnamefont
  {Wang}},\ }\bibfield  {title} {\bibinfo {title} {{Computing by Robust
  Transience: How the Fronto-Parietal Network Performs Sequential,
  Category-Based Decisions}},\ }\href
  {https://doi.org/10.1016/j.neuron.2017.03.002} {\bibfield  {journal}
  {\bibinfo  {journal} {Neuron}\ }\textbf {\bibinfo {volume} {93}},\ \bibinfo
  {pages} {1504} (\bibinfo {year} {2017})}\BibitemShut {NoStop}%
\bibitem [{\citenamefont {Kurikawa}\ \emph {et~al.}(2018)\citenamefont
  {Kurikawa}, \citenamefont {Haga}, \citenamefont {Handa}, \citenamefont
  {Harukuni},\ and\ \citenamefont {Fukai}}]{Kurikawa2018}%
  \BibitemOpen
  \bibfield  {author} {\bibinfo {author} {\bibfnamefont {T.}~\bibnamefont
  {Kurikawa}}, \bibinfo {author} {\bibfnamefont {T.}~\bibnamefont {Haga}},
  \bibinfo {author} {\bibfnamefont {T.}~\bibnamefont {Handa}}, \bibinfo
  {author} {\bibfnamefont {R.}~\bibnamefont {Harukuni}},\ and\ \bibinfo
  {author} {\bibfnamefont {T.}~\bibnamefont {Fukai}},\ }\bibfield  {title}
  {\bibinfo {title} {{Neuronal stability in medial frontal cortex sets
  individual variability in decision-making}},\ }\href
  {https://doi.org/10.1038/s41593-018-0263-5} {\bibfield  {journal} {\bibinfo
  {journal} {Nature Neuroscience}\ }\textbf {\bibinfo {volume} {21}},\ \bibinfo
  {pages} {1764} (\bibinfo {year} {2018})}\BibitemShut {NoStop}%
\bibitem [{\citenamefont {Amari}(1977)}]{Amari1977}%
  \BibitemOpen
  \bibfield  {author} {\bibinfo {author} {\bibfnamefont {S.-i.}\ \bibnamefont
  {Amari}},\ }\bibfield  {title} {\bibinfo {title} {{Neural Theory of
  Association and Concept-Formation}},\ }\href@noop {} {\bibfield  {journal}
  {\bibinfo  {journal} {Biological Cybernetics}\ }\textbf {\bibinfo {volume}
  {26}},\ \bibinfo {pages} {175} (\bibinfo {year} {1977})}\BibitemShut
  {NoStop}%
\bibitem [{\citenamefont {Hopfield}(1984)}]{Hopfield1984}%
  \BibitemOpen
  \bibfield  {author} {\bibinfo {author} {\bibfnamefont {J.~J.}\ \bibnamefont
  {Hopfield}},\ }\bibfield  {title} {\bibinfo {title} {{Neurons with graded
  response have collective computational properties like those of two-state
  neurons.}},\ }\href
  {http://www.pubmedcentral.nih.gov/articlerender.fcgi?artid=345226&tool=pmcentrez&rendertype=abstract}
  {\bibfield  {journal} {\bibinfo  {journal} {Proceedings of the National
  Academy of Sciences of the United States of America}\ }\textbf {\bibinfo
  {volume} {81}},\ \bibinfo {pages} {3088} (\bibinfo {year}
  {1984})}\BibitemShut {NoStop}%
\bibitem [{\citenamefont {Amit}\ \emph {et~al.}(1987)\citenamefont {Amit},
  \citenamefont {Gutfreund},\ and\ \citenamefont {Sompolinsky}}]{Amit1987}%
  \BibitemOpen
  \bibfield  {author} {\bibinfo {author} {\bibfnamefont {D.}~\bibnamefont
  {Amit}}, \bibinfo {author} {\bibfnamefont {H.}~\bibnamefont {Gutfreund}},\
  and\ \bibinfo {author} {\bibfnamefont {H.}~\bibnamefont {Sompolinsky}},\
  }\bibfield  {title} {\bibinfo {title} {{Statistical mechanics of neural
  networks near saturation}},\ }\href
  {http://www.sciencedirect.com/science/article/B6WB1-4DF4WSH-9K/2/a699bd9ad5dfedb917f5a619cb63a63b}
  {\bibfield  {journal} {\bibinfo  {journal} {Annals of Physics}\ }\textbf
  {\bibinfo {volume} {173}},\ \bibinfo {pages} {30} (\bibinfo {year}
  {1987})}\BibitemShut {NoStop}%
\bibitem [{\citenamefont {Personnaz}\ \emph {et~al.}(1986)\citenamefont
  {Personnaz}, \citenamefont {Guyon},\ and\ \citenamefont
  {Dreyfus}}]{Personnaz1986}%
  \BibitemOpen
  \bibfield  {author} {\bibinfo {author} {\bibfnamefont {L.}~\bibnamefont
  {Personnaz}}, \bibinfo {author} {\bibfnamefont {I.}~\bibnamefont {Guyon}},\
  and\ \bibinfo {author} {\bibfnamefont {G.}~\bibnamefont {Dreyfus}},\
  }\bibfield  {title} {\bibinfo {title} {{Collective computational properties
  of neural networks: New learning mechanisms}},\ }\href
  {http://pra.aps.org/abstract/PRA/v34/i5/p4217_1} {\bibfield  {journal}
  {\bibinfo  {journal} {Physical Review A}\ }\textbf {\bibinfo {volume} {34}},\
  \bibinfo {pages} {4217} (\bibinfo {year} {1986})}\BibitemShut {NoStop}%
\bibitem [{\citenamefont {Kanter}\ and\ \citenamefont
  {Sompolinsky}(1987)}]{Kanter1987}%
  \BibitemOpen
  \bibfield  {author} {\bibinfo {author} {\bibfnamefont {I.}~\bibnamefont
  {Kanter}}\ and\ \bibinfo {author} {\bibfnamefont {H.}~\bibnamefont
  {Sompolinsky}},\ }\bibfield  {title} {\bibinfo {title} {{Associative recall
  of memory without errors}},\ }\href@noop {} {\bibfield  {journal} {\bibinfo
  {journal} {Physical Review A}\ }\textbf {\bibinfo {volume} {35}},\ \bibinfo
  {pages} {380} (\bibinfo {year} {1987})}\BibitemShut {NoStop}%
\bibitem [{\citenamefont {Diederich}\ and\ \citenamefont
  {Opper}(1987)}]{Diederich1987}%
  \BibitemOpen
  \bibfield  {author} {\bibinfo {author} {\bibfnamefont {S.}~\bibnamefont
  {Diederich}}\ and\ \bibinfo {author} {\bibfnamefont {M.}~\bibnamefont
  {Opper}},\ }\bibfield  {title} {\bibinfo {title} {{Learning of correlated
  patterns in spin-glass networks by local learning rules}},\ }\href
  {http://adsabs.harvard.edu/abs/1987PhRvL..58..949D} {\bibfield  {journal}
  {\bibinfo  {journal} {Physical review letters}\ }\textbf {\bibinfo {volume}
  {58}},\ \bibinfo {pages} {949} (\bibinfo {year} {1987})}\BibitemShut
  {NoStop}%
\bibitem [{Note1()}]{Note1}%
  \BibitemOpen
  \bibinfo {note} {When $\beta \rightarrow \infty $, $a=1$ and $b=0$ for $0 <
  \gamma <2 $, while $a=0$ and $b=1$ for $2 < \gamma $, which are obtained from
  Eqs. \ref {eq:ab1} and \ref {eq:ab2}}\BibitemShut {NoStop}%
\bibitem [{Note2()}]{Note2}%
  \BibitemOpen
  \bibinfo {note} {$x^{\protect \text {fp}}$ and the chaotic attractor coexist
  for some range of parameters and do not otherwise. Even when they coexist,
  the neural states converge into the chaotic attractor for almost all initial
  states, as shown in Figs. \ref {fig:recall}C, \ref {fig:depend_a}C and S1C.
  Thus, the type of recall is independent of their coexistence.}\BibitemShut
  {Stop}%
\bibitem [{\citenamefont {Kurikawa}\ and\ \citenamefont
  {Kaneko}(2012)}]{Kurikawa2012}%
  \BibitemOpen
  \bibfield  {author} {\bibinfo {author} {\bibfnamefont {T.}~\bibnamefont
  {Kurikawa}}\ and\ \bibinfo {author} {\bibfnamefont {K.}~\bibnamefont
  {Kaneko}},\ }\bibfield  {title} {\bibinfo {title} {{Associative memory model
  with spontaneous neural activity}},\ }\href
  {https://doi.org/10.1209/0295-5075/98/48002} {\bibfield  {journal} {\bibinfo
  {journal} {EPL (Europhysics Letters)}\ }\textbf {\bibinfo {volume} {98}},\
  \bibinfo {pages} {48002} (\bibinfo {year} {2012})}\BibitemShut {NoStop}%
\bibitem [{\citenamefont {Rajan}\ \emph {et~al.}(2010)\citenamefont {Rajan},
  \citenamefont {Abbott},\ and\ \citenamefont {Sompolinsky}}]{Rajan2010}%
  \BibitemOpen
  \bibfield  {author} {\bibinfo {author} {\bibfnamefont {K.}~\bibnamefont
  {Rajan}}, \bibinfo {author} {\bibfnamefont {L.~F.}\ \bibnamefont {Abbott}},\
  and\ \bibinfo {author} {\bibfnamefont {H.}~\bibnamefont {Sompolinsky}},\
  }\bibfield  {title} {\bibinfo {title} {{Stimulus-dependent suppression of
  chaos in recurrent neural networks}},\ }\href
  {https://doi.org/10.1103/PhysRevE.82.011903} {\bibfield  {journal} {\bibinfo
  {journal} {Physical Review E}\ }\textbf {\bibinfo {volume} {82}},\ \bibinfo
  {pages} {011903} (\bibinfo {year} {2010})}\BibitemShut {NoStop}%
\bibitem [{\citenamefont {Toyoizumi}\ and\ \citenamefont
  {Abbott}(2011)}]{Toyoizumi2011}%
  \BibitemOpen
  \bibfield  {author} {\bibinfo {author} {\bibfnamefont {T.}~\bibnamefont
  {Toyoizumi}}\ and\ \bibinfo {author} {\bibfnamefont {L.~F.}\ \bibnamefont
  {Abbott}},\ }\bibfield  {title} {\bibinfo {title} {{Beyond the edge of chaos:
  Amplification and temporal integration by recurrent networks in the chaotic
  regime}},\ }\href {https://doi.org/10.1103/PhysRevE.84.051908} {\bibfield
  {journal} {\bibinfo  {journal} {Physical Review E}\ }\textbf {\bibinfo
  {volume} {84}},\ \bibinfo {pages} {051908} (\bibinfo {year}
  {2011})}\BibitemShut {NoStop}%
\bibitem [{\citenamefont {Shotaro}\ and\ \citenamefont
  {Toshio}(2023)}]{Takasu2023}%
  \BibitemOpen
  \bibfield  {author} {\bibinfo {author} {\bibfnamefont {T.}~\bibnamefont
  {Shotaro}}\ and\ \bibinfo {author} {\bibfnamefont {A.}~\bibnamefont
  {Toshio}},\ }\bibfield  {title} {\bibinfo {title} {{Suppression of chaos in a
  partially driven recurrent neural network}},\ }\href@noop {} {\bibfield
  {journal} {\bibinfo  {journal} {arXiv}\ } (\bibinfo {year} {2023})},\ \Eprint
  {https://arxiv.org/abs/2306.00900} {2306.00900} \BibitemShut {NoStop}%
\bibitem [{\citenamefont {Mastrogiuseppe}\ and\ \citenamefont
  {Ostojic}(2018)}]{Mastrogiuseppe2018}%
  \BibitemOpen
  \bibfield  {author} {\bibinfo {author} {\bibfnamefont {F.}~\bibnamefont
  {Mastrogiuseppe}}\ and\ \bibinfo {author} {\bibfnamefont {S.}~\bibnamefont
  {Ostojic}},\ }\bibfield  {title} {\bibinfo {title} {{Linking Connectivity,
  Dynamics, and Computations in Low-Rank Recurrent Neural Networks}},\ }\href
  {https://doi.org/10.1016/j.neuron.2018.07.003} {\bibfield  {journal}
  {\bibinfo  {journal} {Neuron}\ }\textbf {\bibinfo {volume} {99}},\ \bibinfo
  {pages} {1} (\bibinfo {year} {2018})}\BibitemShut {NoStop}%
\bibitem [{\citenamefont {Schuessler}\ \emph {et~al.}(2020)\citenamefont
  {Schuessler}, \citenamefont {Dubreuil}, \citenamefont {Mastrogiuseppe},
  \citenamefont {Ostojic},\ and\ \citenamefont {Barak}}]{Schuessler2020}%
  \BibitemOpen
  \bibfield  {author} {\bibinfo {author} {\bibfnamefont {F.}~\bibnamefont
  {Schuessler}}, \bibinfo {author} {\bibfnamefont {A.}~\bibnamefont
  {Dubreuil}}, \bibinfo {author} {\bibfnamefont {F.}~\bibnamefont
  {Mastrogiuseppe}}, \bibinfo {author} {\bibfnamefont {S.}~\bibnamefont
  {Ostojic}},\ and\ \bibinfo {author} {\bibfnamefont {O.}~\bibnamefont
  {Barak}},\ }\bibfield  {title} {\bibinfo {title} {{Dynamics of random
  recurrent networks with correlated low-rank structure}},\ }\href
  {https://doi.org/10.1103/PhysRevResearch.2.013111} {\bibfield  {journal}
  {\bibinfo  {journal} {Physical Review Research}\ }\textbf {\bibinfo {volume}
  {2}},\ \bibinfo {pages} {013111} (\bibinfo {year} {2020})}\BibitemShut
  {NoStop}%
\bibitem [{\citenamefont {Kurikawa}\ and\ \citenamefont
  {Kaneko}(2013)}]{Kurikawa2013}%
  \BibitemOpen
  \bibfield  {author} {\bibinfo {author} {\bibfnamefont {T.}~\bibnamefont
  {Kurikawa}}\ and\ \bibinfo {author} {\bibfnamefont {K.}~\bibnamefont
  {Kaneko}},\ }\bibfield  {title} {\bibinfo {title} {{Embedding responses in
  spontaneous neural activity shaped through sequential learning.}},\ }\href
  {https://doi.org/10.1371/journal.pcbi.1002943} {\bibfield  {journal}
  {\bibinfo  {journal} {PLoS computational biology}\ }\textbf {\bibinfo
  {volume} {9}},\ \bibinfo {pages} {e1002943} (\bibinfo {year}
  {2013})}\BibitemShut {NoStop}%
\bibitem [{\citenamefont {Kurikawa}\ \emph {et~al.}(2020)\citenamefont
  {Kurikawa}, \citenamefont {Barak},\ and\ \citenamefont
  {Kaneko}}]{Kurikawa2020}%
  \BibitemOpen
  \bibfield  {author} {\bibinfo {author} {\bibfnamefont {T.}~\bibnamefont
  {Kurikawa}}, \bibinfo {author} {\bibfnamefont {O.}~\bibnamefont {Barak}},\
  and\ \bibinfo {author} {\bibfnamefont {K.}~\bibnamefont {Kaneko}},\
  }\bibfield  {title} {\bibinfo {title} {{Repeated sequential learning
  increases memory capacity via effective decorrelation in a recurrent neural
  network}},\ }\href {https://doi.org/10.1103/PhysRevResearch.2.023307}
  {\bibfield  {journal} {\bibinfo  {journal} {Physical Review Research}\
  }\textbf {\bibinfo {volume} {2}},\ \bibinfo {pages} {023307} (\bibinfo {year}
  {2020})}\BibitemShut {NoStop}%
\bibitem [{\citenamefont {Churchland}\ \emph {et~al.}(2010)\citenamefont
  {Churchland}, \citenamefont {Yu}, \citenamefont {Cunningham}, \citenamefont
  {Sugrue}, \citenamefont {Cohen}, \citenamefont {Corrado}, \citenamefont
  {Newsome}, \citenamefont {Clark}, \citenamefont {Hosseini}, \citenamefont
  {Scott}, \citenamefont {Bradley}, \citenamefont {Smith}, \citenamefont
  {Kohn}, \citenamefont {Movshon}, \citenamefont {Armstrong}, \citenamefont
  {Moore}, \citenamefont {Chang}, \citenamefont {Snyder}, \citenamefont
  {Lisberger}, \citenamefont {Priebe}, \citenamefont {Finn}, \citenamefont
  {Ferster}, \citenamefont {Ryu}, \citenamefont {Santhanam}, \citenamefont
  {Sahani},\ and\ \citenamefont {Shenoy}}]{Churchland2010}%
  \BibitemOpen
  \bibfield  {author} {\bibinfo {author} {\bibfnamefont {M.~M.}\ \bibnamefont
  {Churchland}}, \bibinfo {author} {\bibfnamefont {B.~M.}\ \bibnamefont {Yu}},
  \bibinfo {author} {\bibfnamefont {J.~P.}\ \bibnamefont {Cunningham}},
  \bibinfo {author} {\bibfnamefont {L.~P.}\ \bibnamefont {Sugrue}}, \bibinfo
  {author} {\bibfnamefont {M.~R.}\ \bibnamefont {Cohen}}, \bibinfo {author}
  {\bibfnamefont {G.~S.}\ \bibnamefont {Corrado}}, \bibinfo {author}
  {\bibfnamefont {W.~T.}\ \bibnamefont {Newsome}}, \bibinfo {author}
  {\bibfnamefont {A.~M.}\ \bibnamefont {Clark}}, \bibinfo {author}
  {\bibfnamefont {P.}~\bibnamefont {Hosseini}}, \bibinfo {author}
  {\bibfnamefont {B.~B.}\ \bibnamefont {Scott}}, \bibinfo {author}
  {\bibfnamefont {D.~C.}\ \bibnamefont {Bradley}}, \bibinfo {author}
  {\bibfnamefont {M.~a.}\ \bibnamefont {Smith}}, \bibinfo {author}
  {\bibfnamefont {A.}~\bibnamefont {Kohn}}, \bibinfo {author} {\bibfnamefont
  {J.~A.}\ \bibnamefont {Movshon}}, \bibinfo {author} {\bibfnamefont {K.~M.}\
  \bibnamefont {Armstrong}}, \bibinfo {author} {\bibfnamefont {T.}~\bibnamefont
  {Moore}}, \bibinfo {author} {\bibfnamefont {S.~W.}\ \bibnamefont {Chang}},
  \bibinfo {author} {\bibfnamefont {L.~H.}\ \bibnamefont {Snyder}}, \bibinfo
  {author} {\bibfnamefont {S.~G.}\ \bibnamefont {Lisberger}}, \bibinfo {author}
  {\bibfnamefont {N.~J.}\ \bibnamefont {Priebe}}, \bibinfo {author}
  {\bibfnamefont {I.~M.}\ \bibnamefont {Finn}}, \bibinfo {author}
  {\bibfnamefont {D.}~\bibnamefont {Ferster}}, \bibinfo {author} {\bibfnamefont
  {S.~I.}\ \bibnamefont {Ryu}}, \bibinfo {author} {\bibfnamefont
  {G.}~\bibnamefont {Santhanam}}, \bibinfo {author} {\bibfnamefont
  {M.}~\bibnamefont {Sahani}},\ and\ \bibinfo {author} {\bibfnamefont {K.~V.}\
  \bibnamefont {Shenoy}},\ }\bibfield  {title} {\bibinfo {title} {{Stimulus
  onset quenches neural variability: a widespread cortical phenomenon.}},\
  }\href {https://doi.org/10.1038/nn.2501} {\bibfield  {journal} {\bibinfo
  {journal} {Nature neuroscience}\ }\textbf {\bibinfo {volume} {13}},\ \bibinfo
  {pages} {369} (\bibinfo {year} {2010})}\BibitemShut {NoStop}%
\bibitem [{\citenamefont {Niessing}\ and\ \citenamefont
  {Friedrich}(2010)}]{Niessing2010}%
  \BibitemOpen
  \bibfield  {author} {\bibinfo {author} {\bibfnamefont {J.~J.}\ \bibnamefont
  {Niessing}}\ and\ \bibinfo {author} {\bibfnamefont {R.~W.}\ \bibnamefont
  {Friedrich}},\ }\bibfield  {title} {\bibinfo {title} {{Olfactory pattern
  classification by discrete neuronal network states}},\ }\href
  {https://doi.org/10.1038/nature08961} {\bibfield  {journal} {\bibinfo
  {journal} {Nature}\ }\textbf {\bibinfo {volume} {465}},\ \bibinfo {pages}
  {47} (\bibinfo {year} {2010})}\BibitemShut {NoStop}%
\bibitem [{\citenamefont {Maass}\ \emph {et~al.}(2002)\citenamefont {Maass},
  \citenamefont {Natschl{\"{a}}ger},\ and\ \citenamefont
  {Markram}}]{Maass2002}%
  \BibitemOpen
  \bibfield  {author} {\bibinfo {author} {\bibfnamefont {W.}~\bibnamefont
  {Maass}}, \bibinfo {author} {\bibfnamefont {T.}~\bibnamefont
  {Natschl{\"{a}}ger}},\ and\ \bibinfo {author} {\bibfnamefont
  {H.}~\bibnamefont {Markram}},\ }\bibfield  {title} {\bibinfo {title}
  {{Real-Time Computing Without Stable States: A New Framework for Neural
  Computation Based on Perturbations}},\ }\href
  {http://dx.doi.org/10.1162/089976602760407955} {\bibfield  {journal}
  {\bibinfo  {journal} {Neural Computation}\ }\textbf {\bibinfo {volume}
  {14}},\ \bibinfo {pages} {2531} (\bibinfo {year} {2002})}\BibitemShut
  {NoStop}%
\bibitem [{\citenamefont {Jaeger}\ and\ \citenamefont
  {Haas}(2004)}]{Jaeger2004}%
  \BibitemOpen
  \bibfield  {author} {\bibinfo {author} {\bibfnamefont {H.}~\bibnamefont
  {Jaeger}}\ and\ \bibinfo {author} {\bibfnamefont {H.}~\bibnamefont {Haas}},\
  }\bibfield  {title} {\bibinfo {title} {{Harnessing Nonlinearity: Predicting
  Chaotic Systems and Saving Energy in Wireless Communication}},\ }\href
  {https://doi.org/10.1126/science.1091277} {\bibfield  {journal} {\bibinfo
  {journal} {Science}\ }\textbf {\bibinfo {volume} {304}},\ \bibinfo {pages}
  {78} (\bibinfo {year} {2004})}\BibitemShut {NoStop}%
\bibitem [{\citenamefont {Bertschinger}\ and\ \citenamefont
  {Natschl{\"{a}}ger}(2004)}]{Bertschinger2004}%
  \BibitemOpen
  \bibfield  {author} {\bibinfo {author} {\bibfnamefont {N.}~\bibnamefont
  {Bertschinger}}\ and\ \bibinfo {author} {\bibfnamefont {T.}~\bibnamefont
  {Natschl{\"{a}}ger}},\ }\bibfield  {title} {\bibinfo {title} {{Real-Time
  Computation at the Edge of Chaos in Recurrent Neural Networks}},\ }\href
  {https://doi.org/10.1162} {\bibfield  {journal} {\bibinfo  {journal} {Neural
  computation}\ }\textbf {\bibinfo {volume} {1436}},\ \bibinfo {pages} {1413}
  (\bibinfo {year} {2004})}\BibitemShut {NoStop}%
\bibitem [{\citenamefont {Legenstein}\ and\ \citenamefont
  {Maass}(2007)}]{Legenstein2007}%
  \BibitemOpen
  \bibfield  {author} {\bibinfo {author} {\bibfnamefont {R.}~\bibnamefont
  {Legenstein}}\ and\ \bibinfo {author} {\bibfnamefont {W.}~\bibnamefont
  {Maass}},\ }\bibfield  {title} {\bibinfo {title} {{Edge of chaos and
  prediction of computational performance for neural circuit models.}},\ }\href
  {https://doi.org/10.1016/j.neunet.2007.04.017} {\bibfield  {journal}
  {\bibinfo  {journal} {Neural Networks}\ }\textbf {\bibinfo {volume} {20}},\
  \bibinfo {pages} {323} (\bibinfo {year} {2007})}\BibitemShut {NoStop}%
\bibitem [{\citenamefont {Tsuda}(2001)}]{Tsuda2001}%
  \BibitemOpen
  \bibfield  {author} {\bibinfo {author} {\bibfnamefont {I.}~\bibnamefont
  {Tsuda}},\ }\bibfield  {title} {\bibinfo {title} {Toward an interpretation of
  dynamic neural activity in terms of chaotic dynamical systems},\ }\href
  {http://journals.cambridge.org/action/displayAbstract?fromPage=online&aid=117259}
  {\bibfield  {journal} {\bibinfo  {journal} {Behavioral and Brain Sciences}\
  }\textbf {\bibinfo {volume} {24}},\ \bibinfo {pages} {793} (\bibinfo {year}
  {2001})}\BibitemShut {NoStop}%
\bibitem [{\citenamefont {Fusi}\ \emph {et~al.}(2016)\citenamefont {Fusi},
  \citenamefont {Miller},\ and\ \citenamefont {Rigotti}}]{Fusi2016}%
  \BibitemOpen
  \bibfield  {author} {\bibinfo {author} {\bibfnamefont {S.}~\bibnamefont
  {Fusi}}, \bibinfo {author} {\bibfnamefont {E.~K.}\ \bibnamefont {Miller}},\
  and\ \bibinfo {author} {\bibfnamefont {M.}~\bibnamefont {Rigotti}},\
  }\bibfield  {title} {\bibinfo {title} {{Why neurons mix: high dimensionality
  for higher cognition}},\ }\href {https://doi.org/10.1016/j.conb.2016.01.010}
  {\bibfield  {journal} {\bibinfo  {journal} {Current Opinion in Neurobiology}\
  }\textbf {\bibinfo {volume} {37}},\ \bibinfo {pages} {66} (\bibinfo {year}
  {2016})}\BibitemShut {NoStop}%
\bibitem [{\citenamefont {Terada}\ and\ \citenamefont
  {Toyoizumi}()}]{Terada2023}%
  \BibitemOpen
  \bibfield  {author} {\bibinfo {author} {\bibfnamefont {Y.}~\bibnamefont
  {Terada}}\ and\ \bibinfo {author} {\bibfnamefont {T.}~\bibnamefont
  {Toyoizumi}},\ }\bibfield  {title} {\bibinfo {title} {{Chaotic neural
  dynamics facilitate probabilistic computations through sampling}},\
  }\href@noop {} {\bibinfo  {journal} {bioRxiv}\ ,\ \bibinfo {pages}
  {2023.05.04.539470}}\BibitemShut {NoStop}%
\bibitem [{\citenamefont {Petermann}\ \emph {et~al.}(2009)\citenamefont
  {Petermann}, \citenamefont {Thiagarajan}, \citenamefont {Lebedev},
  \citenamefont {Nicolelis}, \citenamefont {Chialvo},\ and\ \citenamefont
  {Plenz}}]{Petermann2009}%
  \BibitemOpen
\bibfield  {journal} {  }\bibfield  {author} {\bibinfo {author} {\bibfnamefont
  {T.}~\bibnamefont {Petermann}}, \bibinfo {author} {\bibfnamefont {T.~C.}\
  \bibnamefont {Thiagarajan}}, \bibinfo {author} {\bibfnamefont {M.~A.}\
  \bibnamefont {Lebedev}}, \bibinfo {author} {\bibfnamefont {M.~A.~L.}\
  \bibnamefont {Nicolelis}}, \bibinfo {author} {\bibfnamefont {D.~R.}\
  \bibnamefont {Chialvo}},\ and\ \bibinfo {author} {\bibfnamefont
  {D.}~\bibnamefont {Plenz}},\ }\bibfield  {title} {\bibinfo {title}
  {{Spontaneous cortical activity in awake monkeys composed of neuronal
  avalanches.}},\ }\href {https://doi.org/10.1073/pnas.0904089106} {\bibfield
  {journal} {\bibinfo  {journal} {Proceedings of the National Academy of
  Sciences of the United States of America}\ }\textbf {\bibinfo {volume}
  {106}},\ \bibinfo {pages} {15921} (\bibinfo {year} {2009})}\BibitemShut
  {NoStop}%
\bibitem [{\citenamefont {Ku{\'{s}}mierz}\ \emph {et~al.}(2020)\citenamefont
  {Ku{\'{s}}mierz}, \citenamefont {Ogawa},\ and\ \citenamefont
  {Toyoizumi}}]{Kusmierz2020}%
  \BibitemOpen
  \bibfield  {author} {\bibinfo {author} {\bibfnamefont {{\L}.}~\bibnamefont
  {Ku{\'{s}}mierz}}, \bibinfo {author} {\bibfnamefont {S.}~\bibnamefont
  {Ogawa}},\ and\ \bibinfo {author} {\bibfnamefont {T.}~\bibnamefont
  {Toyoizumi}},\ }\bibfield  {title} {\bibinfo {title} {{Edge of Chaos and
  Avalanches in Neural Networks with Heavy-Tailed Synaptic Weight
  Distribution}},\ }\href {https://doi.org/10.1103/PhysRevLett.125.028101}
  {\bibfield  {journal} {\bibinfo  {journal} {Physical Review Letters}\
  }\textbf {\bibinfo {volume} {125}},\ \bibinfo {pages} {028101} (\bibinfo
  {year} {2020})}\BibitemShut {NoStop}%
\bibitem [{\citenamefont {Sussillo}\ and\ \citenamefont
  {Abbott}(2009)}]{Sussillo2009}%
  \BibitemOpen
  \bibfield  {author} {\bibinfo {author} {\bibfnamefont {D.}~\bibnamefont
  {Sussillo}}\ and\ \bibinfo {author} {\bibfnamefont {L.~F.}\ \bibnamefont
  {Abbott}},\ }\bibfield  {title} {\bibinfo {title} {{Generating coherent
  patterns of activity from chaotic neural networks.}},\ }\href
  {https://doi.org/10.1016/j.neuron.2009.07.018} {\bibfield  {journal}
  {\bibinfo  {journal} {Neuron}\ }\textbf {\bibinfo {volume} {63}},\ \bibinfo
  {pages} {544} (\bibinfo {year} {2009})}\BibitemShut {NoStop}%
\bibitem [{\citenamefont {Orhan}\ and\ \citenamefont {Ma}(2019)}]{Orhan2019}%
  \BibitemOpen
  \bibfield  {author} {\bibinfo {author} {\bibfnamefont {A.~E.}\ \bibnamefont
  {Orhan}}\ and\ \bibinfo {author} {\bibfnamefont {W.~J.}\ \bibnamefont {Ma}},\
  }\bibfield  {title} {\bibinfo {title} {{A diverse range of factors affect the
  nature of neural representations underlying short-term memory}},\ }\href
  {https://doi.org/10.1038/s41593-018-0314-y} {\bibfield  {journal} {\bibinfo
  {journal} {Nature Neuroscience}\ }\textbf {\bibinfo {volume} {22}},\ \bibinfo
  {pages} {275} (\bibinfo {year} {2019})}\BibitemShut {NoStop}%
\end{thebibliography}%


\pagebreak
\widetext
\newpage
\begin{center}
\textbf{\large Supplemental Figures: Solvable Neural Network Model for Input-Output Associations: Optimal Recall at the Onset of Chaos}
\end{center}
\setcounter{equation}{0}
\setcounter{figure}{0}
\setcounter{table}{0}
\setcounter{page}{1}
\makeatletter
\renewcommand{\theequation}{S\arabic{equation}}
\renewcommand{\thefigure}{S\arabic{figure}}
\renewcommand{\bibnumfmt}[1]{[S#1]}
\renewcommand{\citenumfont}[1]{S#1}


\begin{figure*}[p]
  \begin{center}
    \includegraphics[width=150mm]{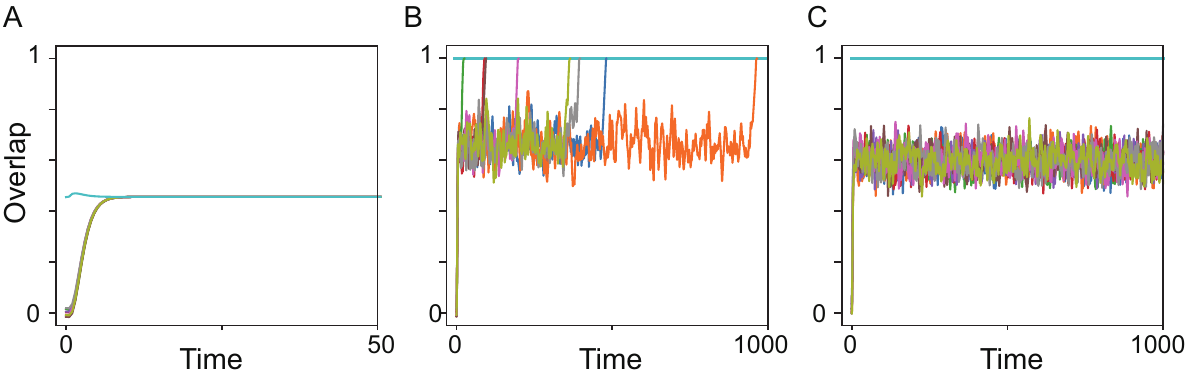}
    \caption { 
    Dynamics of neural activity are shown with their overlap with the target for $\beta = 0.8,3.8$ and $8$ in panels A,B and C, respectively, for $\gamma=1$.
    The value of $\alpha$ is fixed at $0.38$.
    Different colored lines represent trials starting from different initial states; one from $x^{\text{fp}}$ (in cyan) and the others from states that are uniform-randomly chosen from $(-1,1)^N$.
    In A, corresponding to the recall (i), all trajectories converge into $x^{\text{fp}}$ rapidly.
    In B, corresponding to the recall (ii), although all trajectories converge into $x^{\text{fp}}$, the transient time before convergence is drastically longer.
    In C, corresponding to the recall (iii), $x^{\text{fp}}$ is still stable, but the trajectories from random initial states do not escape the chaotic behavior and the chaotic attractor emerges.
        }
    \label{fig:dyn_depn_b}
  \end{center}
\end{figure*}

\begin{figure*}[bp]
  \begin{center}
    \includegraphics[width=150mm]{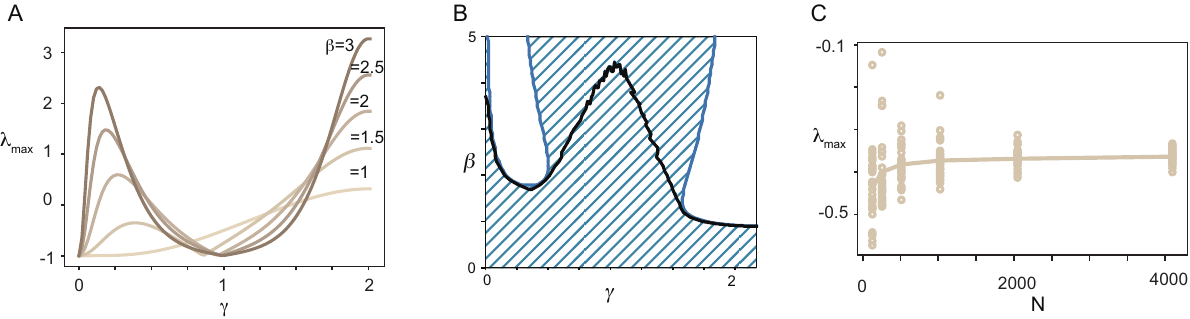}
    \caption { 
    The stability of $x^{\text{fp}}$ is analyzed for varying $\beta$ and $\gamma$, while $\alpha$ is fixed at $0.38$.
     A) Maximum eigenvalue $\lambda_{\text{max}}$ of the Jacobian matrix of $F(x)$ in Eq.(1) at $x^{\text{fp}}$ is plotted for varying $\gamma$ ($N=2048$).
     B) Phase diagram of the stability of $x^{\text{fp}}$ against $\gamma$ and $\beta$ is shown.
     The blue curve  represents the line at which $\lambda_{\text{max}} = 0$, while the black line shows the boundary at which the chaotic attractor loses its stability given in Fig. \ref{fig:recall}B for reference.
In the blue-shaded area, $x^{\text{fp}}$ is stable, whereas $x^{\text{fp}}$ is unstable in the white area.
Either $x^{\text{fp}}$ attractor or the chaotic attractor exists for any parameter space.     
     C) Dependence of $\lambda_{\text{max}}$ on $N$ is plotted. 
     $\lambda_{\text{max}}$ of 10 realizations of networks are computed for $\beta=1.5$ and $\gamma=0.4$ and are plotted as 10 circles.
     The solid line is the value averaged over 10 networks.
     The variability of $\lambda_{\text{max}}$ over different networks reduces as $N$ increases, implying that $\lambda_{\text{max}}$ is determined independent of network realizations and, consequently, the boundary in panel B is also determined independent of the realization of networks, if $N$ is sufficiently large.
        }
    \label{fig:stablity_xfp}
  \end{center}
\end{figure*}

\begin{figure*}[bp]
  \begin{center}
    \includegraphics[width=150mm]{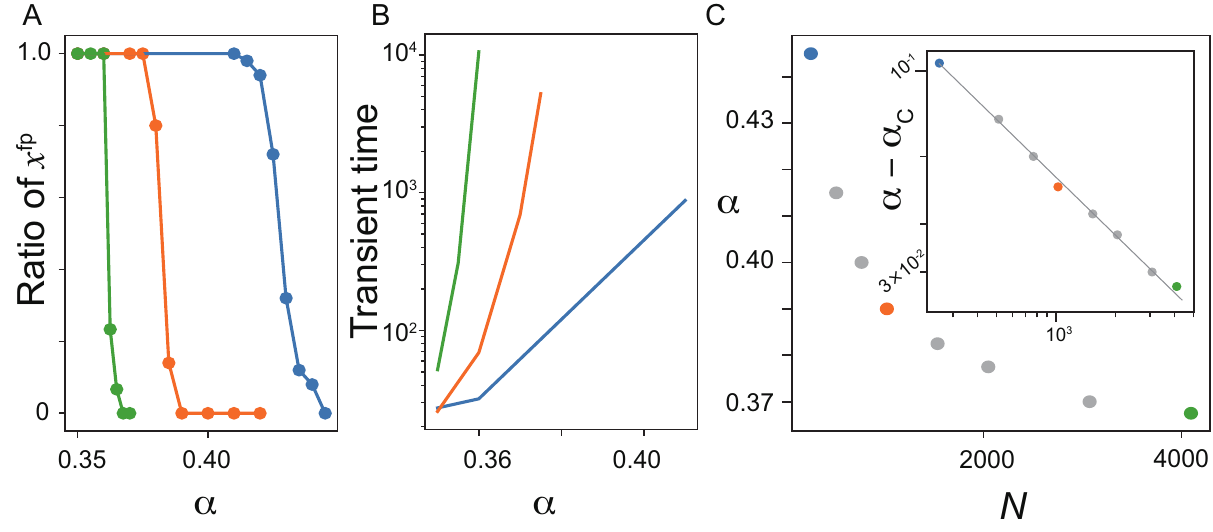}
    \caption {
    Dependence of the memory capacity $\alpha_C$ on $N$.      
The ratio of trajectories converging into $x^{\text{fp}}$ is plotted  with the  increase in $\alpha$ for different $N$.
We numerically computed the ratio by averaging over ten initial states in each of the ten realizations of networks for each $\alpha$.
The ratio that is equal to unity indicates the recall (ii), whereas that equal to zero means that all trajectories converge into the chaotic attractor and consequently the recall (iii).
For identifying $\alpha_C (\infty)$, we analyzed the neural dynamics for quite high $\beta$, here $\beta=32$.
Here, $\gamma$ is set at 1, respectively.
Different colors (blue, orange, and green) represent different $N=256,1024$, and  $4096$, respectively, as same as in B and C.
B)  The transient time before convergence into $x^{\text{fp}}$.
As shown in Fig. \ref{fig:dyn_depn_b}, the transient time becomes much longer as $\alpha$ approaches the transition point $\alpha_C$ from the recall (ii) to (iii). 
We calculated the transition time with the increase in $\alpha$ for different $N$ and plotted it in B.
The transient time diverges quickly when $\alpha$ approaches the transition point $\alpha_C$.
C)  Dependence of $\alpha_C$ on $N$.
To estimate the behavior of $\alpha_C$ at $N \rightarrow \infty$, we calculated $\alpha_C (N)$ with varying $N$.
The data are fitted by $aN^{1/2} + \alpha_C (\infty)$ as shown in the inset, where we derived $\alpha_C (\infty) = 0.340$ ($a=1.67)$.
        }
    \label{fig:alpha_c}
  \end{center}
\end{figure*}

\end{document}